\documentclass{ws-ijmpcs}

\usepackage{url,overcite}
\usepackage{epsfig,amssymb,amsmath,psfrag,subfigure,rotate,slashed,float}
\newcommand{\insertfig}[2]{\mbox{\epsfxsize=#1cm \epsfbox{#2.eps}}}
\newcommand{\bit}[1]{\mbox{\boldmath$#1$}}
\newcommand{\ft}[2]{{\textstyle\frac{#1}{#2}}}
\def\trimmarks{}

\begin{document}

\markboth{Yao Ji, A.V. Belitsky}
{On equations of motion in twist-four evolution}

%
\catchline{}{}{}{}{}
%

\title{On equations of motion in twist-four evolution}

\author{Yao Ji, A.V. Belitsky}

\address{Department of Physics, Arizona State University, Tempe, AZ 85287-1504, USA}

\maketitle


\begin{abstract}
Explicit diagrammatic calculation of evolution equations for high-twist correlation functions is a challenge already at one-loop order in QCD coupling. The main 
complication being quite involved mixing pattern of the so-called non-quasipartonic operators. Recently, this task was completed in the literature for twist-four 
nonsinglet sector. Presently, we elaborate on a particular component of renormalization corresponding to the mixing of gauge-invariant operators with QCD 
equations of motion. These provide an intrinsic contribution to evolution equations yielding total result in agreement with earlier computations that bypassed 
explicit analysis of Feynman graphs.
\end{abstract}

\ccode{PACS numbers: 12.38.Cy, 12.39.St, 25.30.Fj}

\section{Introduction}

Construction of renormalization group evolution equations for twist-four operators represents quite a formidable task even at leading order of QCD perturbative series. 
The main obstacle being, that compared to, say, the twist-three sector \cite{Bukhvostov:1983te} where the entire operator basis can be built from the so-called 
quasipartonic operators \cite{BFLK}, it is no longer the case for higher twists. The reason for this is that the QCD equations of motions and relations based on 
Lorentz symmetry \cite{Bukhvostov:1983te} are not sufficient to eliminate non-quasipartonic operators from the set. As it was established more than three decades 
ago, the quasipartonic operators are distinguished by their property that one can use for them on-shell elementary states in evaluation of corresponding evolution 
kernels and the resulting renormalization group operator takes a pair-wise form at one-loop, conserving the number of fields in correlation functions in question 
\cite{BFLK}. This virtue ceases to be the case when one goes outside of the quasipartonic states. Here off-shell nature of the particles involved mixes states 
with different number of fields. However, since the quasipartonic operators possess the maximal number of fields per given twist, the mixing is unidirectional, only 
transition of non-quasipartonic operators with smaller number of fields is possible into operators with the same number of fields or larger. For instance, for twist-four
operators a generic form of the evolution equations admits the form  
\begin{align}
\frac{d}{d \ln \mu}
\left(
\begin{array}{c}
O_2 \\
O_3 \\
O_4
\end{array}
\right)
=
\left(
\begin{array}{ccc}
K_{2 \to 2} & K_{2 \to 3} & K_{2 \to 4} \\
0 & K_{3 \to 3} & K_{3 \to 4} \\
0 & 0 & K_{4 \to 4}
\end{array}
\right)
\left(
\begin{array}{c}
O_2 \\
O_3 \\
O_4
\end{array}
\right)
\, .
\end{align}
Here $O_N$ is an operator with $N$ elementary QCD fields, with $N=4$ standing for the quasipartionic operator in conventional classification. As we alluded to above,
the mixing matrix takes a triangular form such that quasipartonic operators $O_4$ form an autonomous sector. The latter was studied in full detail decades ago 
\cite{BFLK,BukFro87,Twist3,BraExact98,BelExact98}. Thus the goal of other studies was to fill the gap by analyzing $K_{2 \to n}$ and $K_{3 \to n}$ transitions.

Over the years, only partial results were collected in the literature \cite{ShuVai82,Twist4}. To alleviate the complexity of brute-force perturbative calculations, a framework 
\cite{Braun:2009vc} that heavily employs consequences of conformal boosts and Poincar\'e transformations the transverse plane has been designed and used to find 
coordinate-space evolution kernels for twist-four operator basis. Notice that a proper choice allowed one to eliminate $2 \to n$ transitions altogether thus further simplifying 
the computation. In a parallel development, a direct diagrammatic analysis was performed in the momentum-fraction space \cite{Ji:2014eta}. The latter is a Fourier transform 
of the aforementioned coordinate representation and is used in implementation of evolution in data analyses. 

The use of momentum representation and light-cone gauge $A^+ = 0$ in the aforementioned calculation allowed us to reduce one-loop computations to a set of algebraic, 
though, quite involved manipulations. Staying away from kinematical boundaries not a single loop integral had to be performed. To confirm our calculation a Fourier transformation 
was performed and agreement was found with the light-ray calculation \cite{Braun:2009vc}. Though, the analysis of Ref. \cite{Ji:2014eta} was quite detailed, in the present note 
we will put an emphasis on a most subtle aspect of that calculation, namely, the use of QCD equations of motion for transitions preserving or increasing the number of fields in 
operators in question. In this note, we  will limit ourselves to a few explicit examples pertinent to the use of quark and gluon equations of motion in construction of twist-four evolution 
equations.

\section{Quark equation of motion}

\begin{figure}[t]
\begin{center}
\psfrag{i}[bc][bc]{\scriptsize$i$}
\psfrag{i'}[tc][tc]{\scriptsize$i'$}
\psfrag{a}[bc][bc]{\scriptsize$a$}
\psfrag{d}[tc][tc]{\scriptsize$d$}
\psfrag{a'}[tc][tc]{\scriptsize$a'$}
\psfrag{k1}[cl][cl]{\scriptsize$k_1$}
\psfrag{k2}[cl][cl]{\scriptsize$k_2$}
\psfrag{p1}[cl][cl]{\scriptsize$p_1$}
\psfrag{p2}[cl][cl]{\scriptsize$p_2$}
\psfrag{p3}[cl][cl]{\scriptsize$p_3$}
\psfrag{g3}[cc][cl]{\scriptsize$ $}
\psfrag{a}[bc][bc]{\scriptsize$b$}
\psfrag{b}[tc][tc]{\scriptsize$a$}
\psfrag{a'}[tc][tc]{\scriptsize$a'$}
\mbox{
\begin{picture}(0,120)(170,0)
\put(0,0){\insertfig{2.7}{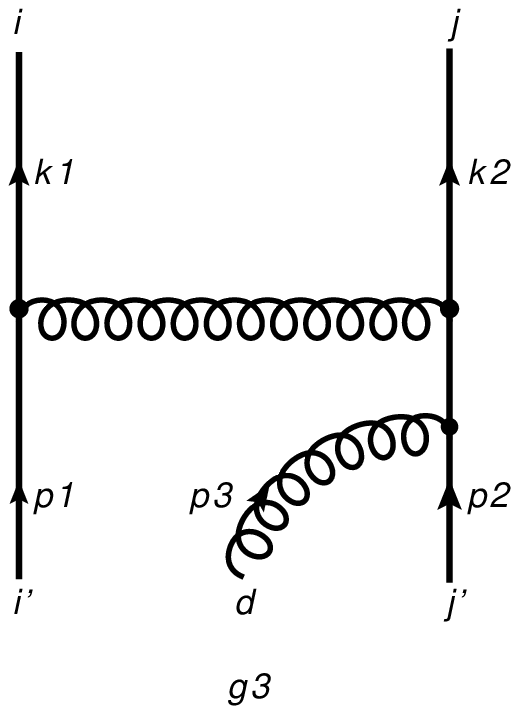}}
\put(130,0){\insertfig{2.7}{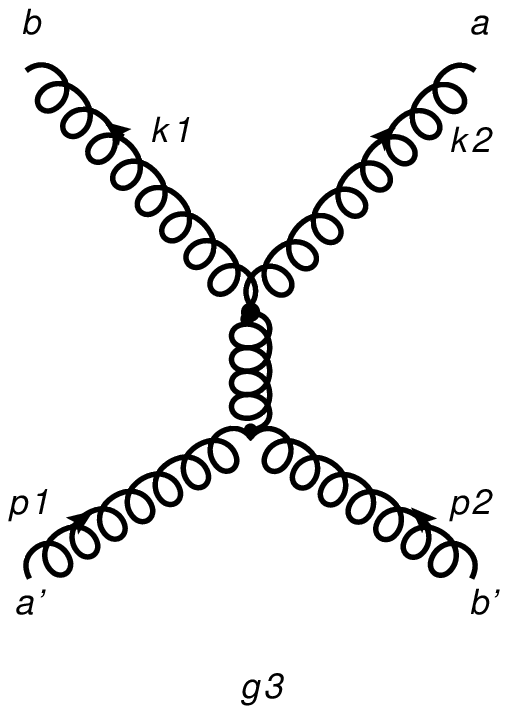}}
\put(260,0){\insertfig{2.8}{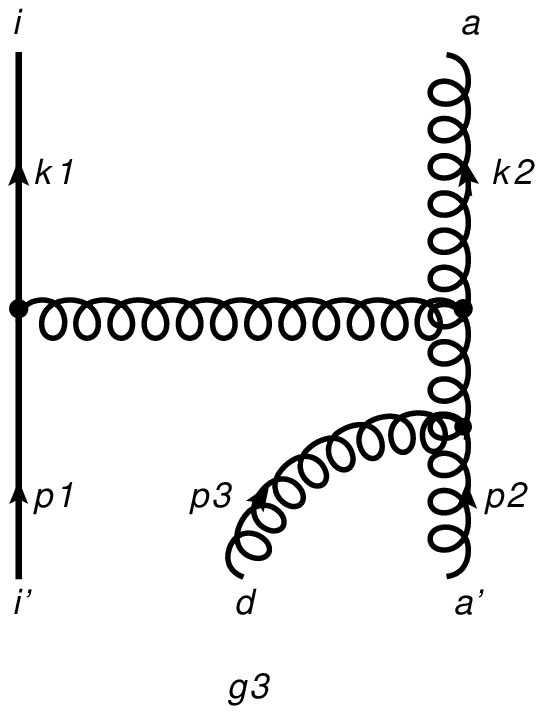}}
\put(30,-5){(a)}
\put(165,-5){(b)}
\put(290,-5){(c)}
\end{picture}
}
\end{center}
\caption{ \label{fig1} Diagrams corresponding to the use of quark and gluon equations of motion in (a) and (b,c), respectively.}
\end{figure}

To start with, let us consider the twist-three sub-block ${\psi}_-^i(x_1)\psi^j_+(x_2)$ built from bad/good components $\psi_{\mp}^i$ of the quark field 
$\psi^i = \psi^i_+ + \psi^i_-$ of a twist-four operator and unravel the contribution from the quark equation of motion in the transition channel involving 
the quasipartonic operator ${\psi}_-^i(x_1)\psi^j_+(x_2) \rightarrow \psi^i_+(y_1)\psi^j(y_2)\bar{f}^d_{++}(y_3)$. This is shown at the diagrammatic level 
in Fig.\ \ref{fig1} (a) and reads in terms of momentum integrals
\begin{align}
\label{QQtoQQG}
&\!
\mathcal{O}_{ij}(x_1,x_2)
=
\int \prod_{i = 1}^3 \frac{d^4p_i}{(2\pi)^4} \delta(p_i^+ - y_i) \prod_{j = 1}^2 \frac{d^4k_j}{(2\pi)^4} \delta(k_j^+ - x_j)
\bar\psi_{i'}(p_1)
i \mathcal{V}^a_{\mu}(p_1,-k_1,-k_3) 
\nonumber\\
&\!\times 
i\mathcal{P}(k_1)
\Gamma^{-+} i\mathcal{P}(-k_2)
i \mathcal{V}^b_{\nu}(p_2+p_3,-k_2,k_3)i\mathcal{P}(-p_2-p_3)i \mathcal{V}^c_{\rho}(-p_2-p_3,p_2,p_3)
\psi_{j'}(p_2)
\nonumber\\
&\qquad
\times A_{\perp}^{d\rho}(p_3)(- i) \Delta^{ab}_{\mu\nu}(k_3)
\, .
\end{align}
Here the operator $\Gamma^{-+}$-matrix, projecting out on appropriate components of four-dimensional Dirac spinors, and the quark-gluon interaction 
vertex are, respectively, 
\begin{align}
\label{fermionpandvertex}
\Gamma^{-+} = \ft{1}{2 \sqrt{2}}\gamma^-\gamma^+(1+\gamma^5)
 \, , \qquad
\mathcal{V}_\mu^a(k_1,k_2,k_3) = g t^a \gamma^\mu (2\pi)^4 \delta ^4(k_1+k_2+k_3)
\end{align}
while the quark and light-cone-gauge gluon propagator read
\begin{align}
\label{gluonp}
\mathcal{P}(k)=\frac{\slashed{k}}{k^2}
\, , \qquad
\Delta^{ab}_{\mu\nu}(k)=\frac{d_{\mu\nu}(k)}{k^2}
\, , \qquad 
d_{\mu\nu}(k)
=
g_{\mu\nu}-\frac{k_{\mu}n_{\nu}+k^{\nu}n_{\mu}}{k_+}
\, .
\end{align}
Denoting the integrand in Eq.~(\ref{QQtoQQG}) by $\mathcal{N}/\mathcal{D}$, we can work out the denominator $\mathcal{D}$ originated from the propagators as 
$\mathcal{D}=k_1^2(p_1-k)^2(p_1+p_2+p_3-k_1)^2(p_2+p_3)^2$. Notice that since all field lines of the quasipartonic quark-gluon operator $\psi^i_+(y_1)\psi^j(y_2)
\bar{f}^d_{++}(y_3)$ are on-shell, one immediately encounters a problem. Namely, the external legs possess collinear momenta $p_i = (p_i^+, 0, \bit{0}_\perp)$ with 
$p_i^+ = y_i$ and thus the propagator $(p_2+p_3)^2$ diverges. To alleviate the problem, one has to properly regularize this. One option is to give a non-vanishing 
minus component $p_i^-$ to particles' momenta. This was done in the past in Refs.\cite{Bukhvostov:1983te}\,. Presently, we follow a different route and use instead the transverse momentum 
as a regulator. Choosing the loop momentum as $k=k_1$, we define $p \equiv p_1+p_2+p_3, q \equiv p_2+p_3$, giving the latter a non-vanishing transverse component
while keeping $q^-=p_2^-+p_3^-=0$. This yields a regularized intermediate propagator $q^2=2q^+q^- - \bit q^2_{\perp}=-\bit q^2_{\perp}$. Expanding the denominator 
$\mathcal{D}$ in the inverse powers of the loop's transverse momentum $\bit{k}_{\perp}$, we find to the lowest few orders
\begin{align}
\frac{1}{\mathcal{D}}&=\frac{1}{\bit{k}_{\perp}^6 \bit{q}_{\perp}^2}\frac{1}{[k^+\beta-1][(k^+-p_1^+)\beta-1][(k^+-p^+)\beta-1]}
\nonumber
\\
&\times\bigg[
1-\frac{2\bit{p}_{1,\perp}\cdot \bit{k}_{\perp}}{\bit{k}^2_{\perp}[(k^+-p_1^+)\beta-1]}-\frac{2 \bit{p}_{1,\perp}\cdot \bit{k}_{\perp}}{\bit{k}^2_{\perp}[(k^+-p^+)\beta-1]}
\bigg]
+
O(1/\bit{k}^8_{\perp})
\, .
\end{align}
Here we kept only terms that induce logarithmic dependence on the ultraviolet cut-off $\mu$ in transverse momentum integrals. These are the renormalization-group
logs that we are resumming. In the above equation, we introduced the $\beta$-variable as a rescaled minus component of the loop momentum $\beta=2k^-/\bit{k}^2_{\perp}$. 
The Dirac algebra in the numerator 
\begin{align}
\mathcal{N}&= - i g^3(t^a)_{ii'} (t^at^d)_{jj'}
\bar{\psi}_{i'}(p_1) [ \gamma^{\mu} \slashed{k} \Gamma^{-+} ( \slashed{k} - \slashed{p}{})\gamma^{\nu} \slashed{q}\slashed{A}_{\perp}^d]  \psi_{j'}(p_2)
d_{\mu\nu} (k-p_1)
\, , 
\end{align}
can be easily performed by means of Sudakov decomposition of all momenta, loop and external. Working this out and performing momentum integrations, with 
$k^+$ momentum simply eliminated by means of the delta-function constraints, $k^-$ momentum (a.k.a.\ $\beta$) generating generalized step-functions \cite{ }
\begin{align}
\label{Generalizedstep}
\vartheta^k_{\alpha_1, \dots, \alpha_n} (x_1, \dots, x_n) 
= 
\int_{-\infty}^{\infty} \frac{d \beta}{2 \pi i} \beta^k \prod_{\ell = 1}^n (x_\ell \beta - 1 + i 0)^{- \alpha_\ell}
\, ,
\end{align}
and, finally, the $\bit{k}_\perp$ one yielding $\ln \mu$ dependence, one immediately arrives at the result
\begin{align}
\mathcal{O}_{ij}(x_1,x_2)&=-\frac{g^3 \ln\mu}{16 \sqrt{2} \pi^2\bit q^2_{\perp}}t^a_{ii'}(t^at^d)_{jj'}
\\
&\times
\int d \mathcal{M}_3 \, 
\bar{\psi}_{i'}(p_1)\bigg[ \frac{\bit q^2_{\perp}\gamma^+\slashed{A}_{\perp}^d(p_3)\vartheta^0_{111}(x_1,x_1-y_1,-x_2)}{x_1-y_1}+ \dots \bigg] \psi_{j'}(p_2)
\, , \nonumber
\end{align} 
where here and below we used the three-particle measure
\begin{align}
d \mathcal{M}_3 \equiv 
\prod_{i=1}^3
\frac{dp_i^-d^2 \bit{p}_{i, \perp}}{(2\pi)^4} \delta \left( \sum_{i=1}^3 y_i - \sum_{j  = 1}^2 x_j \right)
\, .
\end{align}
Since Fig.\ \ref{fig1}~(a) corresponds to the quark equation of motion $\slashed{p}\psi(p) = - g \int d^4 p' \slashed{A}(p')\psi(p - p')$ and $\slashed{q}\slashed{q}=q^2=
-\bit q^2_{\perp}$, we only need to keep terms proportional to $\bit q^2_{\perp}$ thus neglecting everything else denoted by ellipses in the right-hand side of the 
above equation. Effectively, the use of the quark equation of motion can be understood as a contraction of the fermion propagator $i\mathcal{P}(-q)$ into a point. 
Thus, we get the final addendum to the rest of the evolution kernel
\begin{align}
\mathcal{O}_{ij}(x_1,x_2)&=\frac{\sqrt{2} g^3}{32\pi^2}t^a_{ii'}(t^at^d)_{jj'} \ln \mu
\\
&\times
\int d \mathcal{M}_3 \, 
\frac{\vartheta^0_{111}(x_1,x_1-y_1,-x_2)}{x_1-y_1}
\bar{\psi}_{i'}(p_1) \gamma^+\slashed{A}_{\perp}^d(p_3) \psi_{j'}(p_2)
\, , \nonumber
\end{align}
where $\gamma^+\slashed{A}_{\perp}$ provides the correct Dirac structure for the channel. Only after this contribution is accounted for, we total results coincides 
with the one available in the literature.

\section{Gluon equation of motion}

Moving on to the gluon equations of motion, $D_{\mu}F^{\mu\nu}=gj^{\nu}$, the same story applies without significant modifications. It becomes more involved though 
due to Lorentz structures of contributing diagrams. Since the focus of our study was a nonsinglet sector, we are not concerned about gluon-quark transition scenarios.
Therefore, we can safely set the QCD quark current to zero, $j^{\nu}=0$ and simplify the gluon equation of motion to just pure gluodynamics $D_{\mu}F^{\mu\nu}=0$.  
Decomposing it in terms of Sudakov components, we get
\begin{align}
(\partial^+)^2 A^{a-} - {\partial}_{\top}A^a_{\perp} - \partial_{\perp} A^a_{\top} - gf^{abc}(A^b_{\perp} A^c_{\top} + A^b_{\top}A^c_{\perp})
&
=0
\, ,\\
\partial^+(F^{-\perp}\pm F^{-\top})+D^-(F^{+\perp}\pm F^{+\top})\qquad\qquad\qquad\quad
&
\nonumber\\
+
\ft{1}{2}(D^{\perp}\mp D^{\top}) 
\big[\pm(F^{\perp\perp}+F^{\perp\top})\mp(F^{\top\perp}+F^{\top\top})\big]
&=
0
\, ,
\end{align}
in the light cone gauge $A^+=0$. Here, we introduced conventional notations for helicity plus/minus gluon fields
\begin{align}
A_\perp = A^1 + i A^2
\, , \qquad
A_\top = A^1 - i A^2
\, .
\end{align}
The same notation will be used below for holomorphic and antiholomorphic components of any four-vector.

In practice, the use of gluonic equations of motion in Feynman graphs is employed by keeping contributions in numerators that cancel denominators of on-shell propagators, 
in a fashion identical to the one for quarks. Namely, giving these lines a small transverse momentum does the trick. Let us illustrate this methods with a few examples.

We start with a simple example of a two-to-two particle transition $f_{++}^a(x_1)\bar{f}^b_{++}(x_2) \rightarrow f_{++}^{a'}(y_1)\bar{f}_{++}^{b'}(y_2)$ that, due to helicity-conservation, 
possesses an nontrivial annihilation channel as shown in Fig.\ \ref{fig1} (b). This graph yields
\begin{align}
\mathcal{O}^{ab}(x_1,x_2)&
=
\int
\prod_{i = 1}^2
\frac{d^4p_i}{(2\pi)^4} \delta (p_i^+ - y_i)
\prod_{j = 1}^2 \frac{d^4k_j}{(2\pi)^4} \delta(k_j^+-x_j)
\frac{k_1^+k_2^+}{p_1^+p_2^+}
\nonumber\\
&\times
\Gamma^{\rho\mu\nu\sigma}
\, 
\mathcal{V}^{b_1b_2a_1}_{\delta\lambda\alpha}(-k_2,p_1+p_2,-k_1) \mathcal{V}^{a'b'b_3}_{\rho\sigma\tau}(p_1,p_2,-p_1-p_2)
\nonumber\\
&\times
(-i)\Delta_{aa_1}^{\mu\nu}(k_1)(-i)\Delta_{bb_1}^{\nu\delta}(k_2)
(-i)\Delta_{b_2b_3}^{\lambda\tau}(p_1+p_2) A^{a'}_\perp (p_1) A^{b'}_\top (p_2)
\, ,
\end{align}
where we use for convenience a projector in terms of the Dirac matrices
$
\Gamma^{\rho\mu\nu\sigma}
=
\gamma^{\rho}_{\perp} {\gamma}^{\mu}_{\top}{\gamma}^{\phantom{\nu}}_{\perp}\gamma^{\nu}_{\perp} {\gamma}^{\sigma}_{\top}
$
on the transverse holomorphic and antiholomorphic components of the gluon field, and the three-gluon vertex being
\begin{align}
\mathcal{V}^{abc}_{\mu\nu\rho}(k_1,k_2,k_3)
&
=(2\pi)^4\delta^{(4)}(k_1+k_2+k_3)
\\
&
\times
gf^{abc} [ (k_1-k_2)_\rho g_{\mu\nu}+(k_2-k_3)_{\mu}g_{\nu\rho}+(k_3-k_1)_{\nu}g_{\rho\mu}]
\,.
\nonumber
\end{align}
We also defined the transverse four vectors of Dirac matrices $\gamma^{\mu}_{\perp}$, ${\gamma}^{\mu}_\top$ and their (anti)holomorphic combinations  $\gamma_{\perp}$ 
and ${\gamma}_{\top}$,
\begin{align}
\gamma_{\perp}^{\mu}=(0,\gamma^1,\gamma^2,0) \, , \qquad
{\gamma}_{\top}^{\mu}=(0,\gamma^1,-\gamma^2,0)  \, , \qquad \gamma_{\perp}=\gamma^1+i\gamma^2 \, , \qquad {\gamma}_{\top}=\gamma^1-i\gamma^2
\end{align}

After some algebra and integration, we arrive at the expression

\begin{align}
\label{2to2gluoneom}
\mathcal{O}^{ab}(x_1,x_2)
&
=
\frac{g^2\ln\mu}{\pi^2}f^{bca}f^{a'b'c}\gamma_{\perp}
\\
&
\times \int d \mathcal{M}_2 \, A^{a'}_{\perp}(p_1)A^{b'}_{\top}(p_2)
\frac{x_1x_2(x_1-x_2)(y_1-y_2)(\vartheta(-x_2)-\vartheta(x_1))}{y_1y_2(y_1+y_2)^3}
\, , \nonumber
\end{align}
with now two-particle measure
\begin{align}
d \mathcal{M}_2 \equiv 
\prod_{i=1}^2
\frac{dp_i^-d^2 \bit{p}_{i, \perp}}{(2\pi)^4} \delta \left( \sum_{i=1}^2 y_i - \sum_{j  = 1}^2 x_j \right)
\, .
\end{align}
Above, $\vartheta(x)$ is the ordinary step function and $\gamma_{\perp}$ structure is maintained to the end as anticipated. The expression above is obtained 
through the cancellation of the denominator of the on-shell propagator $(p_1+p_2)^2=-(\bit{p}_1+\bit{p}_2)^2_{\perp}$ and ignoring all  terms that fail to 
cancel $(p_1+p_2)^{- 2}$. In fact, the terms that remove the singular dependence on ``divergent'' propagator $(p_1+p_2)^{-2}$ are the only term that survive 
to the end.

Let us now outline the strategy for a two-to-three transitions, shown in Fig.~\ref{fig1} (c). This graphs correspond to the a particular contribution to the 
transition $\tfrac{1}{2}D_{-+}\bar{\psi}^i_+(x_1)\bar{f}^a_{++}(x_2)\rightarrow\bar{\psi}^i_{+}(y_1)\bar{f}^a_{++}(y_2)\bar{f}^d_{++}(y_3)$. Using the vertex
$\Gamma^{\rho \chi} = \ft{\sqrt[4]{2}}{2}\gamma^+ {\gamma}^{\phantom{\rho}}_{\top}{\gamma}^{\rho}_\top{\gamma}^{\chi}_\top$ in four-dimensional notations,
we can write the momentum integrals for the transition in question as 
\begin{align}
\mathcal{O}^{ia}(x_1,x_2)
&
=
\int
\prod_{i = 1}^3
\frac{d^4p_i}{(2\pi)^4} \delta (p_i^+ - y_i) \prod_{j=1}^2 \frac{d^4k_1}{(2\pi)^4} \delta (k_j^+- x_j)
\bar\psi_{i'}(p_1) i \mathcal{V}^{b_1}_{\mu}(p_1,-k_1,-k_3)\nonumber\\
&\times
i\mathcal{P}(k_1) 
\Gamma^{\rho \chi}
{k}_{1\top}A_{\perp}^{d, \alpha} (p_3)
\mathcal{V}^{b_2b_4b_3}_{\nu\sigma\lambda}(k_3,p_2+p_3,-k_2)
\mathcal{V}^{dab_5}_{\alpha\chi\tau}(p_3,p_2,-p_2-p_3)
\nonumber\\
&\times
(-i) \Delta_{b_1b_2}^{\mu\nu}(k_1) (-i) \Delta_{cb_3}^{\rho\sigma}(k_2) (-i) \Delta_{b_4b_5}^{\lambda\tau}(p_2+p_3)
A_\top^{a'} (p_2)
\, , 
\end{align}
where as before $k_{1\top}=k^1-ik^2$ and $k_{1\perp}=k^1+ik^2$. After some algebraic manipulations that cancel the on-shell propagator and 
straightforward integrations over the loop momentum, we end up with the following expression
\begin{align}
\mathcal{O}^{ia}(x_1,x_2)
&
=
-
\frac{\sqrt[4]{2}ig^3}{8\pi^2}t^{b_1}_{ii'}f^{b_1b_4a}f^{da'b_4}
\\
&
\times
\int d \mathcal{M}_3
V (x_1, x_2|y_1, y_2, y_3)
\bar\psi_{i'} (p_1) \gamma^+ \gamma_{\top} A_\top^{a'} (p_2){A}^d_{\top}(p_3)
\, , \nonumber
\end{align}
where the transition kernel stemming from the graph \ref{fig1} (c) reads
\begin{align}
&
V (x_1, x_2|y_1, y_2, y_3)
=
\frac{ x_2 \big[ (y_2+y_3) (y_1+2 (y_2+y_3))-2 y_3 x_2 \big]}{y_2 y_3 (x_1-y_1) (y_2+y_3)}
  \vartheta^0_{12} (x_1-y_1,-x_2)
\nonumber
\\
&+\frac{2 y_1 y_2 (y_2+y_3)-x_2 (y_2-y_3) (y_1+2 (y_2+y_3))+2 x_2^2 (y_2-y_3)}{y_2 y_3 (x_1-y_1) (y_2+y_3)}
 \vartheta^0_{112} (x_1,x_1-y_1,-x_2)
\nonumber\\
&+\frac{ x_1 x_2 (y_3-y_2) \vartheta^0_{111}(x_1,x_1-y_1,-x_2)}{y_2 y_3 (x_1-y_1) (y_2+y_3)} 
-
\frac{ x_1 y_1 x_2  \vartheta^1_{112} (x_1,x_1-y_1,-x_2)}{x_1 y_2 y_3-y_1 y_2 y_3}
\, .
\end{align}

\begin{figure}[t]
\begin{center}
\psfrag{a}[tc][tc]{\scriptsize$a$}
\psfrag{a'}[bc][bc]{\scriptsize$a'$}
\psfrag{d}[tc][tc]{\scriptsize$d$}
\psfrag{i'}[bc][bc]{\scriptsize$i'$}
\psfrag{i}[tc][tc]{\scriptsize$i$}
\psfrag{k1}[cl][cl]{\scriptsize$k_1$}
\psfrag{k2}[cl][cl]{\scriptsize$k_2$}
\psfrag{p1}[cl][cl]{\scriptsize$p_1$}
\psfrag{p2}[cl][cl]{\scriptsize$p_2$}
\psfrag{p3}[cl][cl]{\scriptsize$p_3$}
\psfrag{g9}[cc][cl]{\scriptsize$$}
\psfrag{g15}[cc][cl]{\scriptsize$$}
\psfrag{g16}[cc][cl]{\scriptsize$$}
\mbox{
\begin{picture}(0,120)(170,0)
\put(0,0){\insertfig{2.6}{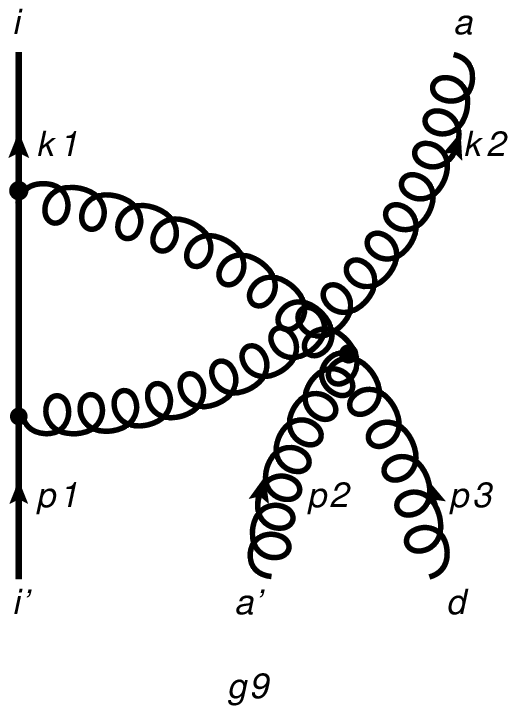}}
\put(130,0){\insertfig{2.6}{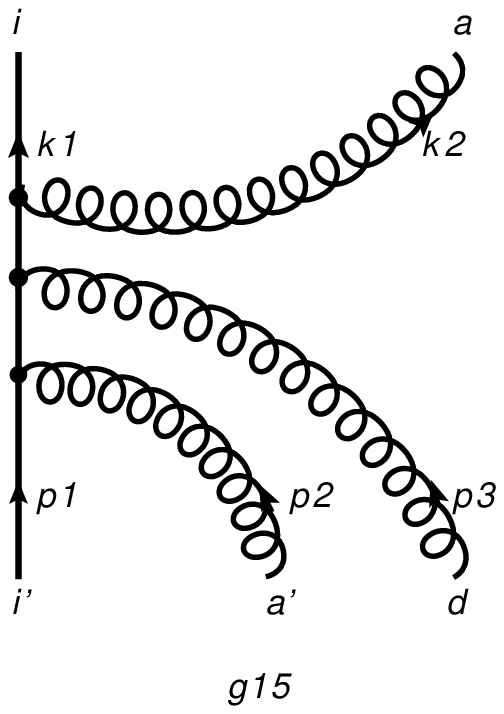}}
\put(260,0){\insertfig{3}{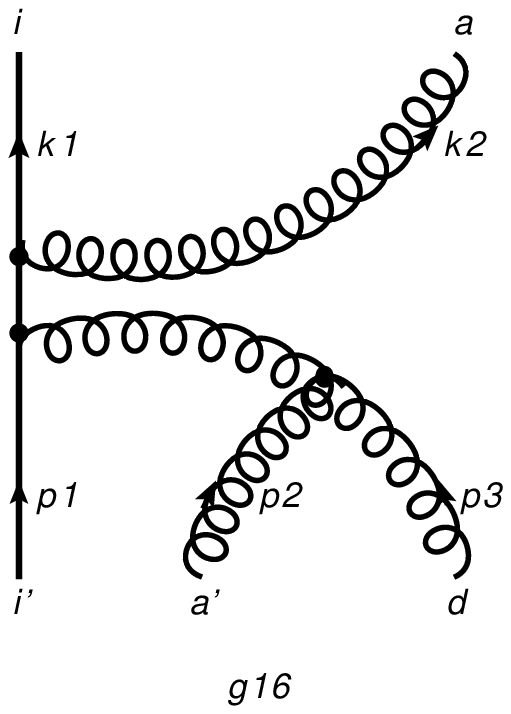}}
\put(30,-5){(a)}
\put(165,-5){(b)}
\put(290,-5){(c)}
\end{picture}
}
\end{center}
\caption{ \label{fig2} In (a), the Feynman graph due to gluon equation of motion in the transition $\tfrac{1}{2}D_{-+}\bar{\psi}^i_+(x_1)f^a_{++}(x_2)
\rightarrow\bar{\psi}^{i'}_{+}(y_1)f^{a'}_{++}(y_2)\bar{f}^d_{++}(y_3)$. In (b) and (c), diagrams corresponding to the double use of equations of motion
for mixing $\tfrac{1}{2}D_{-+}\bar{\psi}^i_+(x_1)f^a_{++}(x_2)\rightarrow\bar{\psi}^{i'}_{+}(y_1)f^{a'}_{++}(y_2)\bar{f}^d_{++}(y_3)$ and 
$\tfrac{1}{2}D_{-+}\bar{\psi}^i_+(x_1)f^a_{++}(x_2)\rightarrow\bar{\psi}^{i'}_{+}(y_1)f^{a'}_{++}(y_2)\bar{f}^d_{++}(y_3)$, respectively.
}
\end{figure}

Let us conclude this section by discussing the use of gluon equations of motion when the field is emitted from an internal off-shell line. The corresponding diagram is
shown in Fig.~\ref{fig2} (a) and reads
\begin{align}
&\!
\mathcal{O}^{ia}(x_1,x_2)
=
\int
\prod_{i = 1}^3
\frac{d^4p_i}{(2\pi)^4} \delta (p_i^+ - y_i) \prod_{j = 1}^2 \frac{d^4k_j}{(2\pi)^4} \delta (k_j^+ - x_j) 
\bar{\psi}_{i'}(p_1)i \mathcal{V}^{b_1}_{\mu}(p_1,-k_2,k_2-p_1)
\nonumber\\
&\ \
\times
i\mathcal{P}(k_1-p_2-p_3) i \mathcal{V}^{b_2}_{\rho}(p_1-k_2,p_2+p_3,-k_1)
 i\mathcal{P}(k_1) \Gamma^{\nu\alpha} k_{1,\top}
\mathcal{V}^{da' b_3}_{\chi\alpha\sigma}(p_3,p_2,-p_2-p_3)
\nonumber\\
&\qquad\times(-i)\Delta_{ab_1}^{\mu\nu}(k_2)(-i)\Delta_{b_2b_3}^{\rho\sigma}(p_2+p_3) A^{a'}_{\perp} (p_2) A^{d,\chi}_{\perp} (p_3)
\, ,
\end{align}
where $\Gamma^{\nu\alpha} = \ft{1}{2 \sqrt{2}} \gamma^+ \gamma_{\top} {\gamma}^{\nu}_{\top} {\gamma}^{\alpha}_{\top}$. After dropping all  terms that leave the
divergent $(p_2+p_3)^2$ factor in the denominator intact, we find
\begin{align}
\mathcal{O}^{ia}(x_1,x_2)
&
=-\frac{\sqrt{2}g^3\ln\mu}{4\pi^2}(t^{a}t^{c})_{ii'}f^{da'c}
\\
&\times
\int d \mathcal{M}_3
V(x_1, x_2| y_1, y_2, y_3) \bar{\psi}_{i'}(p_1) \gamma^+ \gamma_{\top}  {A}^{a'}_{\perp} (p_2) {A}^d_{\top} (p_3)
\, , \nonumber
\end{align}
with the following nontrivial evolution kernel
\begin{align}
&
V(x_1, x_2| y_1, y_2, y_3)
=
-\frac{\left(y_1^2-3 y_1 x_2+2 x_2^2\right) \vartheta^0_{112}(x_1,y_1-x_2,-x_2)}{y_2 y_3 (y_2+y_3)}
\nonumber\\
&
+
\frac{x_1 (y_1-2 x_2) \vartheta^0_{22} (y_1-x_2,-x_2)}{y_2 y_3 (y_2+y_3)}
+
\frac{(2 x_2-y_1) (y_2-y_3) \vartheta^1_{122} (x_1,y_1-x_2,-x_2)}{y_2 y_3}
\nonumber\\
&   
+
\frac{(y_1 (y_3-3 y_2)+2 x_2 (y_2-y_3)) \vartheta^0_{122} (x_1,y_1-x_2,-x_2)}{y_2 y_3 (y_2+y_3)}
\nonumber\\
&
+\frac{4 x_2 (x_2-y_1) (-y_1-y_2-y_3+x_2) \vartheta^2_{122}   (x_1,y_1-x_2,-x_2)}{y_2 (y_2+y_3)}
 \, .
\end{align}

\section{Double equations of motion}

In certain mixing channels, one also has to account for graphs that involve double use of quark and/or gluon equations of motion. In these cases, the method advocated above
works as well. Below, we give a couple of examples to illustrate the point. First, we will discuss the diagram, shown in Fig.\ \ref{fig2} (b), that possesses two quark on-shell 
propagators. Its integral representation reads
\begin{align}
\mathcal{O}^{ia}(x_1,x_2)
&=
\int
\prod_{i = 1}^3
\frac{d^4p_i}{(2\pi)^4} \delta (p_i^+ - y_i)
\prod_{j = 1}^2
\frac{d^4k_i}{(2\pi)^4} \delta (k_j^+ - x_j)
\bar{\psi}^{i'}(p_1)
i \mathcal{V}^{d}_{\rho}(p_1,p_2,-p_1-p_2)
\nonumber\\
&\times
i \mathcal{P}(p_1+p_2) 
i \mathcal{V}^{a'}_{\chi}(p_1+p_2,p_3,-p_1-p_2-p_3)
i \mathcal{P}(p_1+p_2+p_3)
\\
&\times
i \mathcal{V}^{b}_{\mu}(p_1+p_2+p_3,-k_1,-k_2)
i\mathcal{P}(k_1)
\Gamma^{\alpha\chi}_\top
(- i) \Delta_{ab}^{\rho\sigma}(k_2) k_{1\top} {A}_{\perp}^{a'}(p_2)  A^{\rho}_{d\perp} (p_3)
\, , \nonumber
\end{align}
where $\Gamma^{\alpha\chi}_\top = \frac{1}{2 \sqrt{2}} \gamma^+ \gamma_{\top} {\gamma}^{\alpha}_{\top} {\gamma}^{\chi}_{\top}$ projects the quark-gluon operator in
question. Separating the terms with cancelled denominators $(p_1+p_2)^2 (p_1+p_2+p_3)^2$ front the rest, we have the following contribution to the transition
from double quark equation of motion,
\begin{align}
\mathcal{O}^{ia}(x_1,x_2)
&
=
\frac{ig^3\ln\mu}{2 \sqrt{2}\pi^2}(t^{d}t^{a'}t^{a})_{ii'}
\\
&\times
\int d \mathcal{M}_3 \, 
\frac{x_1x_2\vartheta^0_{12}(x_1,-x_2)}{y_2y_3(x_1+x_2)}
\bar{\psi}^{i'}(p_1) \gamma^+ \gamma_{\top} {A}^{a'}_{\perp}(p_2) {A}^d_{\top}(p_3)
\, . \nonumber
\end{align}

Last but not least, let us address the case involving both quark and gluon equations of motion. A representative graph in demonstrated in Fig.~\ref{fig2} (c).
It provided an additive contribution to the following two-to-three transition $\tfrac{1}{2}D_{-+}\bar{\psi}^i_+(x_1)f^a_{++}(x_2)\rightarrow \bar{\psi}^i_+(y_1)f^a_{++}(y_2)
\bar{f}^d_{++}(y_3)$ and gives
\begin{align}
&
\mathcal{O}^{ia}(x_1,x_2)
=
\int \prod_{i =1}^3
\frac{d^4p_i}{(2\pi)^4} \delta(p_i^+-y_i) \prod_{j = 1}^2 \frac{d^4k_j}{(2\pi)^4} \delta(k_j^+-x_j)
{k}_{1,\top} A^{a'\rho}_{\perp}(p_2) A^{d\chi}_{\perp}(p_3)
\nonumber
\\
&
\times\bar{\psi}(p_1) i \mathcal{V}^{b}_{\lambda}(p_1,p_2+p_3,-p_1-p_2-p_3)i\mathcal{P}(p_1+p_2+p_3)i \mathcal{V}^{c}_{\sigma}(-k_1,-k_2,p_1+p_2+p_3)
\nonumber
\\
&\times i\mathcal{P}(k_1)
\Gamma_\top
\mathcal{V}^{b'da'}_{\tau\chi\rho}(-p_2-p_3,p_3,p_2)
(-i) \Delta_{ac}^{\alpha\sigma}(k_2) (-i)\Delta_{bb'}^{\lambda\tau}(p_2+p_3)
\, ,
\end{align}
with $\Gamma_\top = \frac{1}{2\sqrt{2}}\gamma^+ {\gamma}_{\top}$. Projecting on the channel in question, taking care of the symmetrization by swapping  
$A^{a'\rho}_{\perp} \leftrightarrow A^{d\chi}_{\perp}$ and finally dropping terms with uncancelled denominator $(p_2+p_3)^2(p_1+p_2+p_3)^2$, we get
\begin{align}
\mathcal{O}^{ia}(x_1,x_2)
&=
-
\frac{g^3\ln\mu}{8 \sqrt{2} \pi^2}(t^{b}t^{a})_{ii'}f^{ba'd}
\\
&\times
\int d \mathcal{M}_3
\frac{x_1x_2(y_2-y_3)\vartheta^0_{12}(x_1,-x_2)}{(x_1+x_2)y_2y_3(y_2+y_3)}
\bar{\psi}^{i'}(p_1)
\gamma^+ {\gamma}_{\top} {A}_\perp^{a'}(p_2){A}^{d}_\top(p_3)
\, . \nonumber
\end{align}

\section{Conclusions}

In this note, we provided a detailed discussion of a formalism that properly incorporates graphs induced by quark and gluon equations of motion in the renormalization
group evolution of twist-four operators. Neglecting these would yield an incomplete answer for corresponding evolution equations. The method uses regularization of
external off-shell lines by giving them a nonvanishing transverse momentum. This techniques is a bit more advantageous compared to endowing the lines with minus
momentum since the former does not admix momentum fractions in the game. When added to the rest of contributions from Feynman graphs, the complete results was 
found in agreement \cite{Ji:2014eta} with recent analysis of twist-four evolution kernels by means of a different technique \cite{Braun:2009vc}. Our formalism is not tailored 
particularly to the twist-four case and thus can be easily generalized to even higher twists once a complete basis of operators is properly chosen.

\section*{Acknowledgments}

This work was supported by the U.S. National Science Foundation under the grant PHY-1068286.

\bibliographystyle{ws-ijmpcs} 
\bibliography{YJandAB}
\end{document}